\documentstyle[12pt,psfig,draft,lscape,array]{nature-ejb}

\def\simlt{\mathrel{\hbox{\rlap{\hbox{\lower4pt\hbox{$\sim$}}}\hbox{$<$}}}}
\def\simgt{\mathrel{\hbox{\rlap{\hbox{\lower4pt\hbox{$\sim$}}}\hbox{$>$}}}}

\def\ale{\mathrel{\hbox{\rlap{\hbox{\lower4pt\hbox{$\sim$}}}\hbox{$<$}}}}
\def\age{\mathrel{\hbox{\rlap{\hbox{\lower4pt\hbox{$\sim$}}}\hbox{$>$}}}}

\def\ra#1#2#3{#1$^{\rm h}$#2$^{\rm m}$#3$^{\rm s}$}
\def\dec#1#2#3{$#1^\circ#2'#3''$}

\def\spose#1{\hbox to 0pt{#1\hss}}

\begin{document}

\title{\large \bf The Birth of a Relativistic Outflow in the Unusual
$\gamma$-ray Transient Swift\,J164449.3+573451}

\author{
B.~A.~Zauderer\affiliation[1]
     {Harvard-Smithsonian Center for Astrophysics, 60 Garden St., 
      Cambridge, MA 02138, USA},
E.~Berger\affiliationmark[1],
A.~M.~Soderberg\affiliationmark[1],
A.~Loeb\affiliationmark[1],
R.~Narayan\affiliationmark[1],
D.~A.~Frail\affiliation[2]
     {National Radio Astronomy Observatory, P.O. Box 0, Socorro, 
      NM 87801, USA},
G.~R.~Petitpas\affiliationmark[1],
A.~Brunthaler\affiliation[3]
     {Max-Planck-Institut f$\ddot{u}$r Radioastronomie, Auf dem 
      H$\ddot{u}$gel 69, 53121 Bonn, Germany},
R.~Chornock\affiliationmark[1],
J.~M.~Carpenter\affiliation[4]
     {Department of Astronomy, California Institute of Technology, 
      Pasadena, CA 91125, USA},
G.~G.~Pooley\affiliation[5]
     {Mullard Radio Observatory, Cavendish Laboratory, Cambridge 
      UK},
K.~Mooley\affiliationmark[4],
S.~R.~Kulkarni\affiliationmark[4],
R.~Margutti\affiliation[6]
     {INAF Osservatorio Astronomico di Brera, via Bianchi 46, 
      Merate 23807, Italy},
D.~B.~Fox\affiliation[7]
     {Department of Astronomy, Pennsylvania State University, 
      University Park, PA 16802, USA},
E.~Nakar\affiliation[8]
     {School of Physics \& Astronomy, Tel Aviv University, Tel 
      Aviv 69978, Israel},
N.~A.~Patel\affiliationmark[1],
N.~H.~Volgenau\affiliation[9]
     {California Institute of Technology, Owens Valley Radio 
      Observatory, Big Pine, CA 93513, USA},
T.~L.~Culverhouse\affiliationmark[9],
M.~F.~Bietenholz\affiliation[10]
     {Hartebeesthoek Radio Astronomy Observatory, PO Box 443,
      Krugersdorp, 1740 South Africa}$^,$\affiliation[11]
     {Department of Physics and Astronomy, York University, 
      Toronto, Ontario, Canada},
M.~P.~Rupen\affiliationmark[2],
W.~Max-Moerbeck\affiliationmark[4],
A.~C.~S.~Readhead\affiliationmark[4],
J.~Richards\affiliationmark[4],
M.~Shepherd\affiliationmark[4],
S.~Storm\affiliation[12]
     {Department of Astronomy, University of Maryland, College 
      Park, MD 20742},~\&
C.~L.~H.~Hull\affiliation[13]
     {Radio Astronomy Laboratory, University of California, Berkeley, 
      CA 94720}
}

\date{\today}{}
\headertitle{Birth of a Relativistic Outflow in
Swift\,J164449.3+573451}
\mainauthor{Zauderer et al.}

\summary{Active galactic nuclei (AGN), powered by long-term accretion
onto central supermassive black holes, produce\cite{bbr84}
relativistic jets with lifetimes of $\simgt 10^6$ yr that preclude
observations at birth.  Transient accretion onto a supermassive black
hole, for example through the tidal disruption\cite{hil75,ree88} of a
stray star, may therefore offer a unique opportunity to observe and
study the birth of a relativistic jet.  On 2011 March 25, the {\it
Swift} $\gamma$-ray satellite discovered\cite{burrows11} an unusual
transient source (Swift\,J164449.3+573451) potentially
representing\cite{bgm+11,ltc+11} such an event.  Here we present the
discovery of a luminous radio transient associated with
Swift\,J164449.3+573451, and an extensive set of observations spanning
centimeter to millimeter wavelengths and covering the first month of
evolution.  These observations lead to a positional
coincidence\cite{gcn11854} with the nucleus of an inactive galaxy, and
provide direct evidence for a newly-formed relativistic outflow,
launched by transient accretion onto a $10^6$ M$_\odot$ black hole.
While a relativistic outflow was not predicted in this scenario, we
show that the tidal disruption of a star naturally explains the
high-energy properties, radio luminosity, and the inferred rate of
such events.  The weaker beaming in the radio compared to
$\gamma$-rays/X-rays, suggests that radio searches may uncover similar
events out to redshifts of $z\sim 6$.}
\maketitle

\noindent{\it This paper has been submitted to Nature. You are free to 
use the results here for the purpose of your research.  In accordance 
with the editorial policy of Nature, we request that you not discuss 
this result in the press.  If you have any question or need 
clarifications please contact Ashley Zauderer (bzauderer@cfa.harvard.edu) 
or Edo Berger (eberger@cfa.harvard.edu).}

\begin{figure}
\centerline{\psfig{file=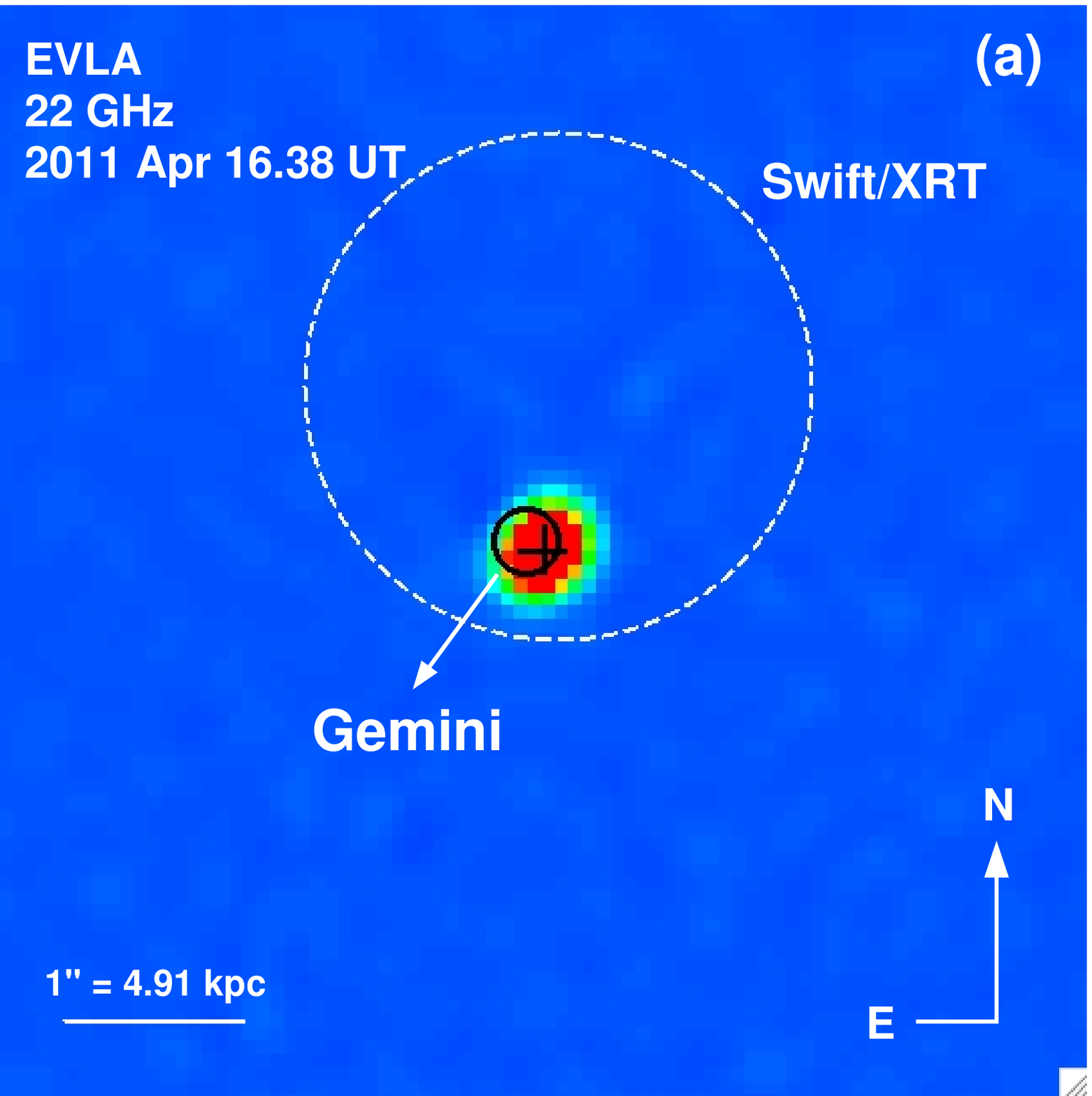,width=2.9in,angle=0}\hfill 
\psfig{file=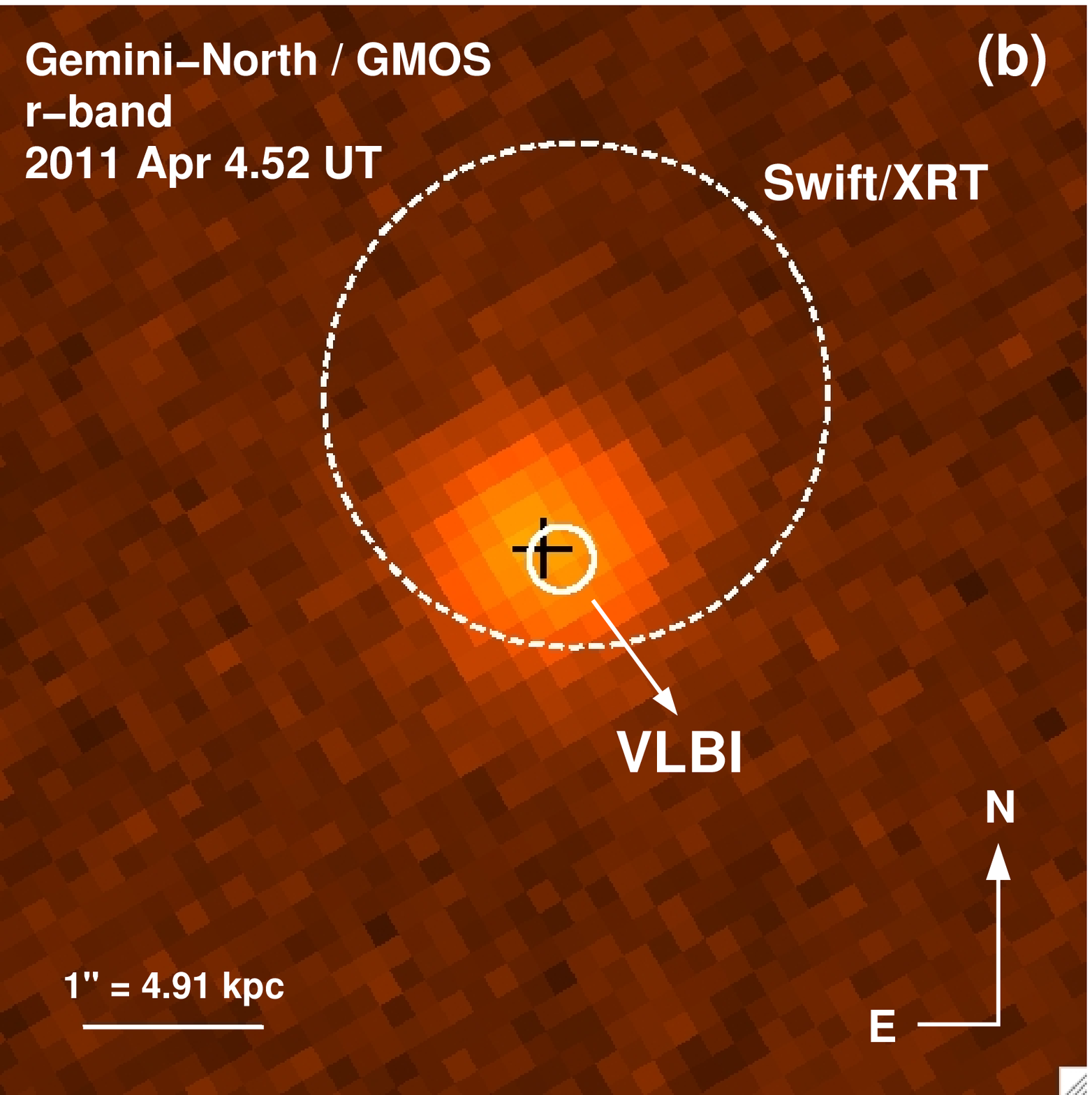,width=2.9in,angle=0}}
\caption[]{\small 
Radio and optical images of Swift\,J164449.3+573451
and its host galaxy reveal a positional alignment between the
transient and the center of the galaxy.  (a) The radio image is from
the EVLA at a frequency of 22 GHz.  The most precise radio position,
from VLBI is $\alpha_{\rm J2000}=$\ra{16}{44}{49.93130}, $\delta_{\rm
J2000}=$\dec{+57}{34}{59.6893} ($\pm 0.1$ mas; see SI).  (b) The
optical $r$-band image was obtained on 2011 April 4.52 UT with the
Gemini-North 8-m telescope, and has been astrometrically aligned to
the Two Micron All Sky Survey (2MASS) catalog using 14 common objects
with a resulting root-mean-square uncertainty of 0.13 arcsec in each
coordinate ($68\%$ confidence level).  The galaxy optical centroid is
located at $\alpha_{\rm J2000}=$\ra{16}{44}{49.942}, $\delta_{\rm
J2000}=$\dec{+57}{34}{59.74} ($\pm 0.01$ arcsec).  The {\it Swift}/XRT
error circle (large white circle), with a radius of 1.4 arcsec ($90\%$
confidence level), contains the galaxy, but cannot be used to locate
the X-ray transient position within it.  On the other hand, the radio
position relative to the astrometric solution of the Gemini image has
an uncertainty of only 0.18 arcsec ($68\%$ confidence level; this
uncertainty is dominated by the astrometric match of the optical image
to the 2MASS catalog, not by the radio position itself) and leads to
an offset of $0.11\pm 0.18$ arcsec, corresponding to a physical scale
of $0.5\pm 0.9$ kpc at $z=0.354$.  The radio transient position is
therefore consistent with an origin in the nucleus of the host galaxy.
(a) The radio centroid is marked by cross-hairs, while the galaxy
optical centroid (with an uncertainty of 0.18 arcsec due to 2MASS
astrometric solution) is marked by the small black circle.  (b) The
galaxy optical centroid is marked by cross-hairs, while the radio
position (with an uncertainty of 0.18 arcsec due to 2MASS astrometric
solution) is marked by the small white circle.
}
\label{fig:image} 
\end{figure}

Upon the discovery\cite{burrows11} of Swift\,J164449.3+573451, and the
identification of a galaxy at a redshift\cite{ltc+11} of $z=0.354$
within the {\it Swift} X-ray localization region ($1.4''$ radius), we
initiated radio observations of the transient on 2011 March 29.36 UT
with the Expanded Very Large Array (EVLA) at a frequency of 5.8 GHz
and discovered an unresolved source with a flux density of 
$310\pm 7$~$\mu$Jy.  Astrometric matching demonstrated that the radio 
source coincides with the galaxy nucleus (Figure~\ref{fig:image}),
subsequently confirmed\cite{ltc+11} with other data.  A follow-up EVLA
observation 0.9~d later revealed that the source brightened to $530\pm
10$~$\mu$Jy, thereby conclusively linking the X-ray/$\gamma$-ray
transient and the galaxy for the first time.  The galaxy exhibits no
evidence\cite{ltc+11} for an active nucleus (see Figure 1 in online
Supplementary Information; SI), and the lack of previous 
$\gamma$-ray/X-ray activity argues\cite{burrows11} against an
AGN flare origin.

\begin{figure}
\centerline{\psfig{file=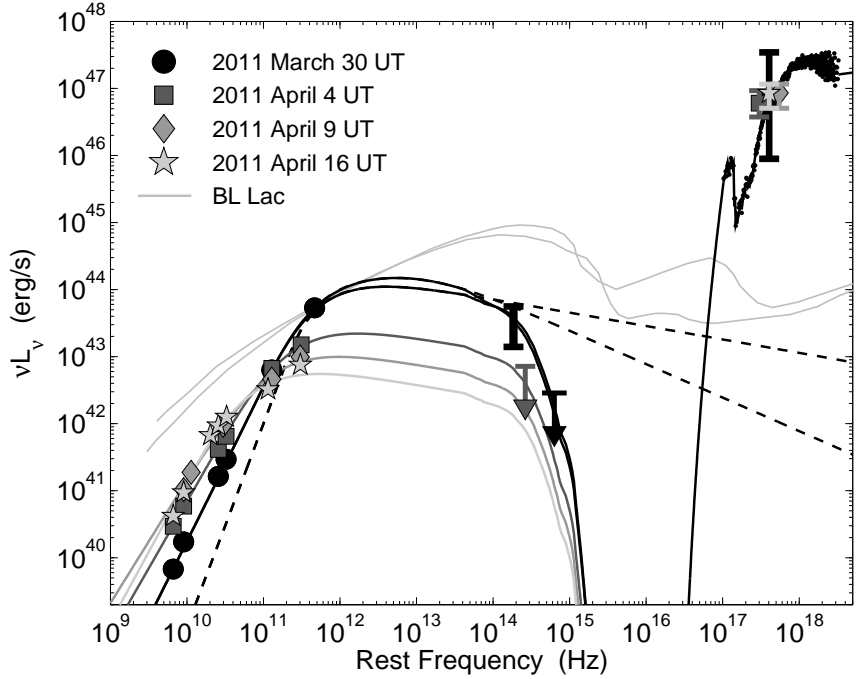,width=4.5in,angle=0}}
\caption[]{\small 
Spectral energy distributions (SEDs) of
Swift\,J164449.3+573451 from radio to X-rays point to synchrotron
emission from a relativistic outflow.  Our radio observations cover
decimeter to millimeter wavelengths at 5, 10, 15, and 22 d after the
initial $\gamma$-ray detection.  The NIR luminosity on March 30
can only be constrained within a factor of five due to the
unknown contribution from the host galaxy; on April 4 the NIR
upper limit is inferred from a {\it Hubble Space Telescope}
image\cite{ltc+11}.  Only an upper bound is available on the optical 
luminosity (black triangle) due to the lack\cite{ltc+11} of variable 
emission.  The flux in the soft X-ray band is highly variable on March
30, but is more quiescent at 10, 15, and 22 d (extrema marked by vertical 
bars and mean brightness by solid symbols with points at 10 and 15 d shifted slightly 
in frequency for clarity).  The radio, NIR, and optical data are well 
modeled by an evolving synchrotron spectrum (solid lines) with 
a large rest-frame optical extinction of $A_V\simgt 3$ mag.  The synchrotron 
curves for the March 30 SED are for two values of the synchrotron cooling frequency: $\nu_c\approx
2\times 10^{13}$ Hz (steeper optically thin slope) and $\nu_c\simgt
2\times 10^{18}$ Hz (shallower optically thin slope).  This 
model cannot explain the large X-ray luminosity, which remains
nearly constant while the radio spectrum is evolving strongly.  A
representative model for the X-ray spectrum (data=black dots;
model=black line) includes power-law ($F_\nu\propto\nu^{0.9}$) and
blackbody ($kT\approx 1$ keV) components with significant absorption
($N_H\approx 2\times 10^{22}$ cm$^{-2}$), in agreement with the
large optical extinction.  Shown for comparison is the SED of the
canonical blazar BL Lac in two separate states (varying in peak
frequency and flux of the synchrotron component), normalized to the
luminosity of Swift\,J164449.3+573451 at 345 GHz.  The blazar SED
provides a poor match.
}
\label{fig:sed} 
\end{figure}

We carried out additional observations at multiple frequencies
spanning $1-345$ GHz with several cm- and mm-wave facilities 
(see SI Table 1). The spectral energy distribution (SED) in this 
frequency range on 2011~March~30~UT ($\Delta t\approx 5$~d) is well 
described by a power law with $F_\nu\propto\nu^{1.3\pm 0.1}$ up to 
$F_\nu(345\,{\rm GHz})=35\pm 1$~mJy.  The steep power law requires 
self-absorbed synchrotron emission.  The weak near-infrared (NIR) 
variability\cite{ltc+11} indicates $F_\nu(2.5\,{\rm\mu m})\approx 
0.1$~mJy, while the lack of optical variability leads\cite{ltc+11} 
to an upper bound of $F_\nu(0.64\,{\rm \mu m})\simlt 2$ $\mu$Jy.  
The SED therefore peaks in the millimeter band, with a best-fit 
rest-frame peak frequency and flux density of $\nu_p\approx 6\times 
10^{11}$~Hz and $F_{\nu,p} \approx 80$~mJy, respectively (
Figure~\ref{fig:sed}).  The non-detection of optical variability 
requires significant rest-frame extinction of $A_V\simgt 3$ mag, 
further supporting a nuclear origin. Subsequent SEDs at 
$\Delta t\approx 10$, $15$, and $22$~d exhibit
significant evolution, with $\nu_p\propto t^{-1.3}$ and
$F_{\nu,p}\propto t^{-0.8}$ (Figure~\ref{fig:sed}).

For synchrotron sources there is a well-defined minimum energy,
achieved\cite{rea94} near equipartition between the fractional
energies in the relativistic electrons ($\epsilon_e$) and magnetic
fields ($\epsilon_B$).  This condition defines\cite{che98} the
equipartition radius: $\theta_{\rm eq}= 110\,d_{\rm
L,Mpc}^{-1/19}\,F_{\rm\nu, p,mJy}^{9/19}\,\nu_{\rm p,GHz}^{-1}$
$\mu$as.  From the March~30~UT SED we find $\theta_{\rm eq}\approx 1$
$\mu$as ($r_{\rm eq}\approx 1.5\times 10^{16}$ cm) and hence mildly
relativistic expansion with a Lorentz factor of $\Gamma\approx 2$.  A
more detailed model that accounts\cite{kn09} for relativistic effects
leads to a similar result, $r_{\rm eq}\approx 1\times 10^{16}$ cm and
$\Gamma\approx 1.2$ (see SI).  From the observed SED temporal
evolution we find that the source continues to expand relativistically
with nearly constant velocity (see SI text).  Extrapolating the linear
trend in radius we find a formation epoch in the range 2011 March
23-26 UT, in excellent agreement with the initial $\gamma$-ray
detection on 2011 March 25 UT (see SI Figure 2).  This provides 
independent evidence for a newly-formed relativistic outflow.

\begin{figure}
\centerline{\psfig{file=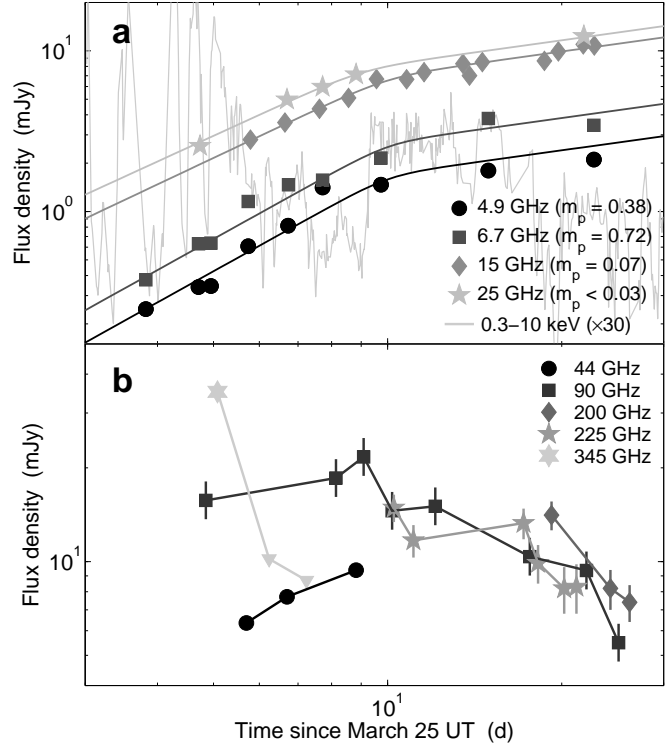,width=3.5in,angle=0}}
\caption[]{\small 
 Radio light curves of Swift\,J164449.3+573451 at
$5-345$ GHz reveal interstellar scintillation.  (a) Light curves at
$5-25$ GHz (error bars are smaller than symbols; see SI).  These data 
are from the EVLA, the AMI Large Array, and the OVRO
40-m telescope. The lines are broken power law fits to the $5-25$ GHz
light curves, using March 25 as the initial time.  The low
frequency light curves exhibit significant interstellar scintillation,
with the strongest modulation at 6.7 GHz.  To calculate the expected
interstellar scintillation we use the NE2001 Galactic Free Electron
Density Model\cite{cl02}.  For the line of sight to Swift\,J164449.3+573451 
($l=86.7111$, $b=39.4415$) the scattering measure is $2.2\times 10^{-4}$ kpc
m$^{-20/3}$.  With a scattering screen distance of $\sim 1$ kpc the
transition from weak to strong scattering occurs\cite{gn06} at 
$\nu_0\approx 10$ GHz, while the Fresnel scale is
$\theta_{F,0}\approx 1$ $\mu$as (sizes are given as radii).  At
frequencies above $\nu_0$ the modulation index is given by $m_p\propto
(\nu/\nu_0)^{-17/12}\,(\theta_s/\theta_{F,0})^{-7/6}$.  For
frequencies below $\nu_0$ refractive scintillation leads to
$m_p\propto (\nu/\nu_0)^{17/30}\,(\theta_s/\theta_r)^{-7/6}$, where
the refractive scale is $\theta_r=\theta_{F,0}(\nu/\nu_0)^{-11/5}$.
Comparing these results to the observed modulation we infer a size of
$\theta_s\approx 5$ $\mu$as.  Also shown is the {\it Swift} X-ray
light curve\cite{burrows11} binned on a timescale of 15 min and
multiplied by a factor of $1.3\times 10^{10}$ to fit on the same flux
density scale as the radio data.  The strong X-ray variability during
the first 10 d is not accompanied by similar order of magnitude
fluctuations in the radio bands, pointing to a distinct origin for the
radio and X-ray emission.  (b) Light curves at $44-345$ GHz from the
EVLA, CARMA, and the SMA (error bars are one standard deviation).
These frequencies are mainly in the decline phase and therefore
provide information on the peak of the synchrotron spectrum
(Figure~\ref{fig:sed}).  Upper limits at 345 GHz are marked by
triangles.
}
\label{fig:lc} 
\end{figure}

An angular size of a few $\mu$as will inevitably lead to variability
in the low frequency radio emission due to interstellar scintillation,
with the amplitude of modulation depending\cite{gn06} on the ratio of
the source size ($\theta_s$) to the Fresnel scale ($\theta_F$).  For
the line of sight to Swift\,J164449.3+573451 the maximum modulation
($m_p\approx 1$) is expected\cite{cl02} at $\nu_0\approx 10$~GHz, for
$\theta_s\approx\theta_F\approx 1$~$\mu$as.  The observed modulation
inferred from our detailed radio light curves is tens of percent at
$5-7$ GHz and a few percent at $15$~GHz (Figure~\ref{fig:lc}), leading
to a projected radius of $\theta_s\approx 5$ $\mu$as, or $\Gamma
\approx 5$.  This provides independent evidence for a relativistic
outflow.  Our radio observations with Very Long Baseline
Interferometry (VLBI) at a frequency of 22 GHz place an upper bound on
the size of $r\simlt 0.8$ pc, consistent with the synchrotron and
scintillation analyses, and providing an upper bound on the lifetime
of the event of $\simlt 1.7$ yr for an expansion with $\Gamma\approx
2$ (see SI text).

The mean X-ray luminosity during the four radio epochs exceeds the
synchrotron peak by a factor of $\sim 10^3$ and therefore requires a
distinct origin (Figure~\ref{fig:sed}).  One potential mechanism to
generate the large X-ray luminosity is inverse Compton (IC) scattering
of radio synchrotron photons by the relativistic electrons
(synchrotron self-Compton: SSC), but from the relativistic model we
find a predicted SSC X-ray luminosity of only $\approx 2\times
10^{45}$ erg s$^{-1}$ (see SI).  Similarly, although order of
magnitude variations in brightness are seen in the X-rays, our
detailed radio light curves do not reveal coincident variations as
would be expected for SSC (Figure~\ref{fig:lc}).  We therefore
conclude that the X-ray emission is dominated by a distinct, and more
compact emission region, most likely at the base of the outflow.

Having established the birth of a relativistic outflow, coincident
with the nucleus of the host galaxy, we briefly describe a model to
power the outflow through transient accretion\cite{bgm+11} onto a
supermassive black hole (SMBH).  The host galaxy luminosity,
$M_B\approx -18.2$ mag, implies\cite{gh07} a modest SMBH mass of $\sim
10^5-10^6$ M$_\odot$.  The duration of the bright early phase in the
X-ray light curve, $\sim 10^5$ s, coincides with the debris fallback
time for a solar-mass star tidally disrupted at a pericenter distance
$R_p\sim 13 (M_{\rm SMBH}/10^6\,{\rm M}_\odot)^{-5/6}$ Schwarzschild
radii.  The most bound stellar debris is expected to feed the black
hole at an initial rate of\cite{sq09} $\sim (\frac{1}{2}{\rm
M_\odot})/10^5$.  With a radiative efficiency of $\simgt 1\%$ at $\sim
R_p$, this can account for the observed X-ray luminosity.  However,
this luminosity is $\sim 10^3$ times the Eddington limit for a $\sim
10^6$ M$_\odot$ black hole, leading inevitably to a highly collimated
outflow, the origin of the radio-emitting relativistic outflow found
here.

We conclude with several key implications of our results.  First, our
initial estimate of the energy, $E_K\approx 3\times 10^{50}$ erg at
$\Delta t\approx 22$ d (see SI Table 2), corresponds to the Eddington
luminosity of a $10^6$ M$_\odot$ black hole, lending support to the
tidal disruption scenario, and suggesting that the X-ray/$\gamma$-ray
emission is collimated by a factor of $\sim 10^3$.  Long-term radio
monitoring will test this result by providing precise
beaming-independent calorimetry\cite{fwk00,sb11} of the true energy
release.  Continued radio observations will also uniquely probe the
density structure near a previously-dormant supermassive black hole as
the ambient medium is swept up by the relativistic outflow.  
From the existing data we find $n_e$~$\propto$~r$^{-2.4}$ (see SI).
Second,
with continued expansion we expect that VLBA observations will resolve
the radio source on a timescale of $\sim 2$~yr, and directly confirm
the relativistic expansion; from the observed flux density evolution
we predict a peak of a few mJy at several GHz on this timescale,
within the reach of the VLBA.  Third, from the detection of a single
such event in 6 years of {\it Swift} operations we infer a rate of
$\sim 0.1$ Gpc$^{-3}$ yr$^{-1}$, much lower than the
predicted\cite{wm04} tidal disruption rate of $\sim 10^2-10^3$
Gpc$^{-3}$ yr$^{-1}$, or upper limits from current radio
surveys\cite{bow11} of $\simlt 10^3$ Gpc$^{-3}$ yr$^{-1}$.  This
suggests that the properties of Swift\,J164449.3+573451 are
exceedingly rare; if due to jet collimation, the implied beaming
fraction is $\sim 10^{3}$, consistent with the ratio of the observed
X-ray luminosity to the Eddington limit and the radio-inferred energy.
Finally, past searches\cite{kb99,gbm+08,cbk+11} for tidal disruption
events have focused on the expected\cite{sq09} bright optical/UV and
soft X-ray emission, but the large optical extinction and associated
soft X-ray absorption in Swift\,J164449.3+573451 suggest that radio
observations may provide a cleaner signature.  This is particularly
true if the X-ray/$\gamma$-ray emission is beamed by a factor of $\sim
10^3$.  With the EVLA and the Atacama Large Millimeter Array, similar
events are detectable to $z\sim 6$.


\clearpage


\noindent$\bf{Acknowledgements}$  E.B.~is supported in part by funds 
from NASA.  A.~L.~is supported in part by NSF and NASA grants.  
R.M.~acknowledges support from a Swift ASI grant and from the Ministry 
of University and Research of Italy.  E.N.~is partially supported by IRG 
and ISF grants.  The EVLA and VLBA are operated by the NRAO, a facility
of the NSF operated under cooperative agreement by AUI.  The SMA is
a joint project between the SAO and the ASIAA, and is funded by the 
Smithsonian Institution and the Academia Sinica.  CARMA development and 
operations are supported by the NSF under a cooperative 
agreement, and by the Associates of the California Institute of Technology, 
the University of Chicago, and the states of California, Illinois, and Maryland.
The AMI arrays are supported by the University of Cambridge and the STFC.  
This work is partially based on observations with the 100-m telescope of 
the MPIfR at Effelsberg.  This work made use of data supplied by the UK 
Swift Science Data Centre at the University of Leicester.

\hspace{1in}

\noindent$\bf{Author~Contributions}$  A.Z. and E.B. designed and 
coordinated the radio observations and 
analysis among all instruments reported here.  A.Z. and D.A.F. 
performed EVLA observations, data reduction and analysis.  
G.R.P. observed the source with the SMA, and along with N.A.P. 
reduced and analyzed the SMA observations.  CARMA observations 
were set up, reduced and analyzed by A.Z., J.M.C. and S.R.K.  
S.S. and C.L.H.H. obtained the first CARMA observations.  
N.H.V. and T.L.C. facilitated quick response CARMA observations.  
R.C. implemented and analyzed MMT and Gemini optical observations.  
K.M. performed observations with the OVRO 40-m and analyzed results, 
with advice from S.R.K., A.C.S.R., J.R., M.S., and W.M.  G.G.P. 
performed observations with the AMI Large Array and analyzed the 
results.  A.M.S., A.B., M.F.B. and M.P.R. planned observations 
with the VLBA and Effelsberg.  A.B. reduced the VLBI data.  
R.M. analyzed and modeled the X-ray data.  A.L., R.N. and E.N. 
provided the theoretical model for a tidal disruption event.  
The paper was synthesized by A.Z. and E.B. with the primary text 
written by E.B. and portions of the SI written by E.B., A.Z., R.C., 
K.M. and A.B.  D.B.F. provided feedback on the manuscript.   
All authors discussed the results and commented on the manuscript.

\hspace{1in}

\noindent$\bf{Author~Information}$  Correspondence and requests for materials
should be addressed to A.Z. (bzauderer@cfa.harvard.edu) or
E.B. (eberger@cfa.harvard.edu).

\clearpage

\clearpage

\clearpage 

\end{document}


\begin{center}
{\Large SUPPLEMENTARY INFORMATION} \\
\end{center}

\headertitle{Supplementary Information}
\mainauthor{Zauderer et al.}

\section{Archival Radio Non-detections of Swift\,J164449.3+573451}

We inspected the location of Swift\,J164449.3+573451 in archival 1.4
GHz images from the NRAO VLA Sky Survey (NVSS) taken in March 1995,
and the Faint Images of the Radio Sky at Twenty-cm (FIRST) survey
taken in July 1998.  No counterpart is detected to $3\sigma$ limits of
about $1.5$ mJy and 0.45 mJy, respectively.  The non-detections place
an upper bound on the radio luminosity of 
$6\times~10^{30}$~erg~s$^{-1}$~Hz$^{-1}$ (March 1995) and 
$2\times~10^{30}$~erg~s$^{-1}$~Hz$^{-1}$ (July 1998), 
thereby ruling out radio-loud AGN activity
(typically\cite{mpm90} $\simgt~10^{33}$~erg~s$^{-1}$~Hz$^{-1}$).

\section{Expanded Very Large Array Observations}
\label{sec:evla}

We observed Swift\,J164449.3+573451 with the EVLA beginning 0.82~d
after the $\gamma$-ray trigger (see Table~\ref{tab:radio}). We used
the EVLA's new Wideband Interferometric Digital Architecture (WIDAR)
correlator to obtain up to 2~GHz of bandwidth at several frequencies.
At all frequencies, we used 3C286 for bandpass and flux calibration.
Absolute flux calibration is accurate to $\sim$~10$\%$.
At 1.4~GHz, we used J1634+6245 for phase calibration.  For phase
calibration at all other frequencies, we used J1638+5720 and also
included a third calibrator, J1639+5357, at 5.8~GHz.
Polarization calibration was not performed.  The data were
reduced and imaged with the Astronomical Image Processing System
(AIPS) software package.

\section{AMI Large Array observations}
\label{sec:ryle}

We observed with 6 antennas of the AMI Large
Array (Mullard Radio Astronomy Observatory, Cambridge, UK) at 15.4 GHz
with a bandwidth of 3.75~GHz beginning 2.79~d after the $\gamma$-ray
trigger (see Table~\ref{tab:radio}).  The maximum baseline is about
110~m and the angular resolution is 25~arcsec.  Observations ranged in
duration from 45~min to 11~hr.  Observations of the compact source
J1638+5720 were interleaved at intervals of 10 m as a phase reference,
and the flux density scale was established by regular observations of
the calibrators 3C48 and 3C286.  Absolute flux calibration is
accurate to better than 10$\%$.  The AMI array uses linearly-polarized 
feeds and measures Stokes I+Q.  It does not measure any other combination 
of Stokes parameters.  Polarization calibration is not reported here.  

\section{Owens Valley Radio Observatory 40-m Observations}
\label{sec:ovro40}

We observed Swift\,J164449.3+573451 with the OVRO 40-meter telescope
at a frequency of 15~GHz beginning 2.82~d after the $\gamma$-ray
trigger (see Table~\ref{tab:radio}).  The 40-m telescope is equipped
with a dual-beam Dicke-switched receiver with two symmetric, off-axis
beams (each 2.5~arcmin full-width at half-maximum) separated
azimuthally by 12.95~arcmin.  The receiver has a 2.5~GHz
noise-equivalent reception bandwidth.  We used sky switching,
alternating the source between the two beams to reduce atmospheric and
ground pickup, and to account for the non-identical nature of the two
beams.  See Ref.~\pcite{ovro} for further details of the telescope
and receiver system.  The flux scale is derived from observations of 
3C286 using standard spectral models and coefficients\cite{baars}
 with an absolute uncertainty of $\sim$~$5\%$.

We note that daytime measurements of Swift\,J164449.3+573451 lead to
consistently discrepant flux densities, most likely due to increased
thermal emission from the ground.  These discrepant measurements, as
quoted in an initial GCN circular, have been used
elsewhere$^{5}$ to argue for a short variability timescale in
the radio emission, and hence a large Lorentz factor of 
$\Gamma\simgt$~10.  We do not support this conclusion, 
and indeed, as we show in
Section~\ref{sec:synch}, $\Gamma\simgt$~10 appears to lead to a
non-self consistent model.

\section{Combined Array for Research in Millimeter Astronomy
Observations}
\label{sec:carma}

We observed Swift\,J164449.3+573451 with CARMA beginning 1.85 d after
the $\gamma$-ray trigger (see Table~\ref{tab:radio}).  CARMA is a
heterogeneous array comprised of nine 6.1-m antennas and six 10.4-m
antennas.  Observations were taken at a rest frequency of 87.3 and
93.6 GHz with a total bandwidth ranging between 6.8 and 7.8~GHz.  We
used Neptune as our primary flux calibrator, and J1824+568 and
J1638+573 as bandpass and phase calibrators, respectively.  The
overall uncertainty in the absolute flux calibration is $\sim 15\%$.
Data calibration and imaging were done with the MIRIAD software
package.

\section{Submillimeter Array Observations}
\label{sec:sma}

We observed Swift\,J164449.3+573451 with the Submillimeter Array (SMA;
Ref.~\pcite{hml04}) beginning 0.28 d after the burst, followed by
monitoring at $\sim 1.3$ mm and $\sim 0.9$ mm (see
Table~\ref{tab:radio}).  Observations were made using at least seven
of the eight antennas, in a wide range of weather conditions, with
$\tau_{\rm 225}$ ranging from 0.04 to 0.3.  For each observation, the
full 8~GHz (4~GHz in each sideband separated by 10~GHz) were combined
to increase the signal-to-noise ratio.  The data were calibrated using
the MIR software package developed at Caltech and modified for the
SMA, while for imaging and analysis we used MIRIAD.  Gain calibration
was performed using J1642+689, 3C345, and J1849+670.  Absolute flux
calibration was performed using real-time measurements of the system
temperatures, with observations of Neptune to set the scale, accurate
to within $\sim$ 10 $\%$.
Bandpass calibration was done using 3C454.3, J1924-292, and 3C279.

\section{Very Long Baseline Array Observations}
\label{sec:vlba}

We observed Swift\,J164449.3+573451 with the NRAO Very Long Baseline
Array (VLBA) and the 100 m Effelsberg telescopes starting on 2011
April 2.25~UT for a duration of 7~hr at 8.4 and 22~GHz.  The
observations were performed with eight frequency bands of 8~MHz
bandwidth each in dual circular polarization, resulting in a total
data rate of 512 Mbps. We used J1638+5720, located only 0.92$^\circ$
from Swift\,J164449.3+573451, as a phase-referencing source at both
frequencies.  At 22 GHz, sources were switched every 40 seconds, while
at 8.4 GHz we spent 40 seconds on the calibrator and 90 seconds on
target.  A second calibrator, J1657+5705, was also observed for six
minutes at each frequency to check the calibration of the data.  We
also employed\cite{brf05,rmb+09} $\sim 30$~min of geodetic blocks at
the start and end of the observations for atmospheric calibration.  

The data were data were correlated at the VLBA Array Operations Center
in Socorro, New Mexico, and calibrated\cite{kvr+06} using AIPS and
ParselTongue.  We applied the latest values of the Earth's orientation
parameters, and performed zenith delay corrections based on the
results of the geodetic block observations.  Total electron content
maps of the ionosphere were used to correct for ionospheric phase
changes.  Amplitude calibration used system temperature measurements
and standard gain curves.  We performed ``manual phase-calibration''
using the data from one scan of J1638+5720 to remove instrumental
phase offsets among the frequency bands.  We then fringe fitted the
data from J1638+5720.  Since J1638+5720 shows extended structure, we
performed phase self-calibration first, and then amplitude and phase
self-calibration to construct robust models at both frequencies.  The
calibration was transferred to the target source and J1657+5705.
Residual phase calibration errors were removed by performing one round
of phase self-calibration on the target with a solution interval of 30
min. The data were imaged in AIPS using robust weighting (ROBUST=0).

The 8.4 GHz images were restored with a beam of $1.03\times 0.46$~mas
at a position angle of $-29^\circ$.  We achieved a noise level of 30
$\mu$Jy beam$^{-1}$, which is close to the expected thermal noise
limit.  Swift\,J164449.3+573451 was detected with a flux density of
$1.71\pm 0.03$ mJy beam$^{-1}$.  The 22 GHz images were restored with
a beam of $0.44\times 0.24$ mas at a position angle of $2^\circ$.  We
achieved a noise level of 120 $\mu$Jy beam$^{-1}$, which is twice the
expected thermal noise limit.  Swift\,J164449.3+573451 was detected
with a flux density of $4.68\pm 0.12$ mJy~beam$^{-1}$.
Absolute flux calibration is $\sim$~15-20$\%$.

The position of the source at both frequencies was measured from the
purely phased referenced image (without phase self-calibration on the
target).  The formal uncertainties of the positions are 
$\approx 3-8$~$\mu$as.  The true uncertainty is dominated by systematic errors
induced from residual atmospheric effects and an opacity effect
(``core shift'') in the calibrator source. The position difference
between the two frequencies is 61~$\mu$as in right ascension and 
33~$\mu$as in declination.  The absolute position as determined from the
22~GHz image is $\alpha_{\rm J2000}=$\ra{16}{44}{49.93130},
$\delta_{\rm J2000}=$\dec{+57}{34}{59.6893} with an uncertainty of 
0.1~mas ($68\%$ confidence level) in each coordinate.

The limit of 0.25~mas corresponds to $r\simlt 2\times 10^{18}$~cm.
This places an upper bound on the Lorentz factor of $\Gamma\simlt 15$
(for relativistic expansion with $r\approx\Gamma^2ct$ and a start time
of 2011 March 25).  Alternatively, we can use the inferred Lorentz
factor of $\Gamma\sim 2$ to place an upper bound on the lifetime of
the source of $\simlt 1.5$ yr.

\section{Optical Spectroscopy}
\label{sec:opt}

We obtained two spectra of the host galaxy of Swift\,J164449.3+573451.
The first spectrum was a 7200 s observation taken at a mean time of
2011 April 1.34 UT using Hectospec on the 6.5-m MMT and covered the
observed wavelength range $3700-9150$ \AA.  The second one was taken
in nod-and-shuffle mode using GMOS on the 8-m Gemini-North telescope
on 2011 April 4.62 UT (program GN-2011A-Q-4).  The R400 grating was
used to cover the wavelength range $4930-9180$ \AA.

\begin{figure}[h!]
\centerline{\psfig{file=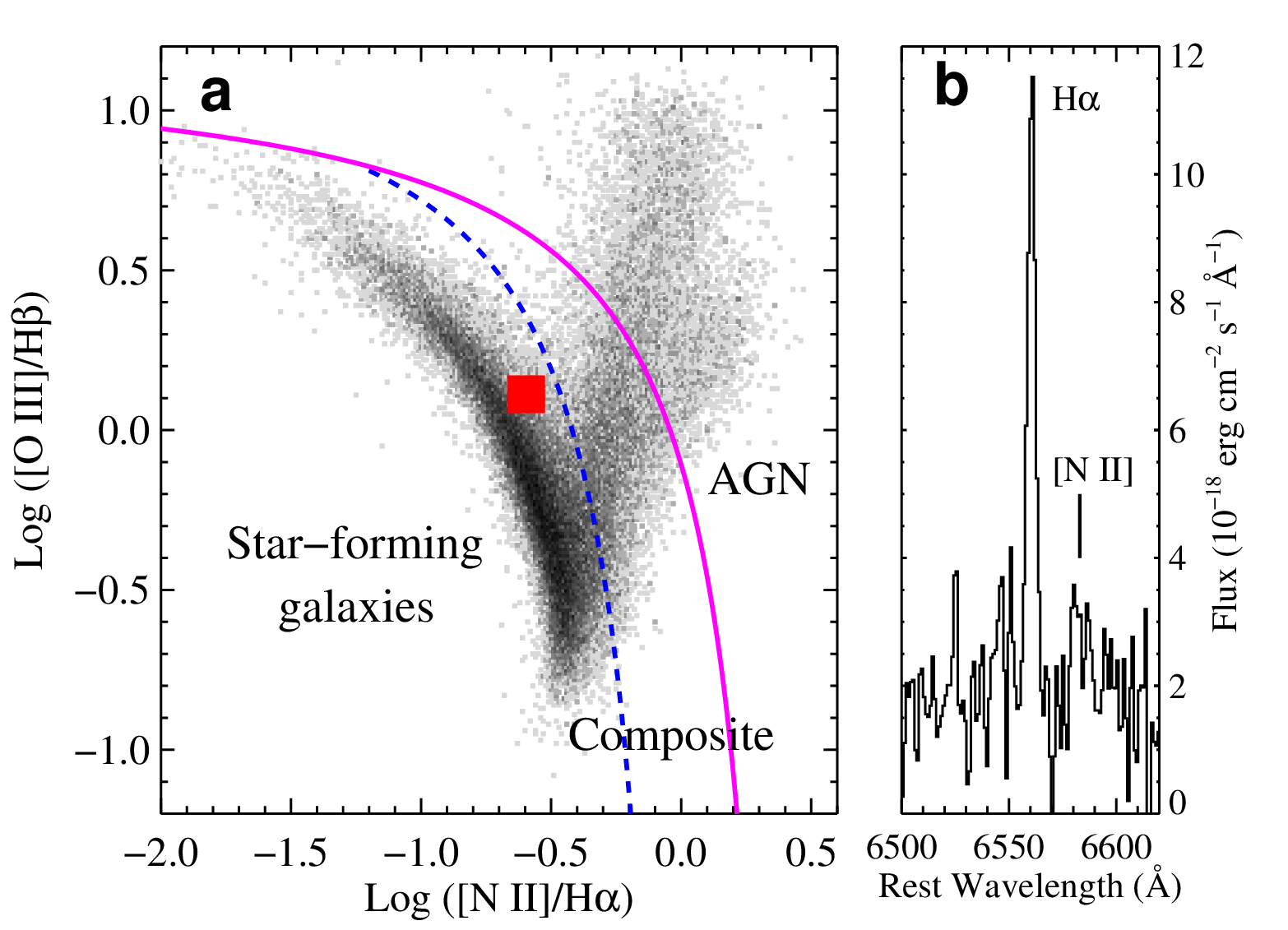,width=6.5in,angle=0}}
\caption[]{\small Optical spectroscopy of the host galaxy of 
Swift\,J164449.3+573451 shows no indication of AGN activity.
(a) Excitation diagram comparing the
observed line ratios of the host galaxy of Swift\,J164449.3+573451 to
a sub-sample of galaxies observed\cite{kht+03,thk+04} in the Sloan
Digital Sky Survey.  The emission lines in objects found to the right
of the magenta line are powered by accretion onto massive black holes,
while those between the blue dashed line and the solid magenta line
show\cite{kgk+06} ``composite'' spectra of star formation plus AGN
activity.  Objects to the left of both lines have emission lines
dominated\cite{kgk+06} by star formation activity.  The host galaxy
(red square) is clearly in the portion of the diagram associated with
star-forming galaxies.  (b) The region around the H$\alpha$
line in the Gemini spectrum.  No broad component is observed as would
be expected for an active Seyfert 1 galaxy.}
\label{fig:host} 
\end{figure}

The spectra are consistent with each other and exhibit the usual
narrow emission lines of a star-forming galaxy ${\rm [OII]}\lambda
3727$, H$\beta$, ${\rm [OIII]}\lambda\lambda 4959,5007$, H$\alpha$) at
a redshift of $z=0.354$.  Neither spectrum shows any evidence for a
broad (${\rm FWHM}\simgt 1000$~km~s$^{-1}$) component to the H$\alpha$
line (SI Figure~\ref{fig:host}b), as would be seen in a
Seyfert 1 galaxy (i.e., a galaxy with an unobscured actively-accreting
central massive black hole).  To examine the possibility that the host
of Swift\,J164449.3+573451 contained a central black hole that was
actively accreting prior to the current outburst, but which was
obscured from our line of sight, we examined the ratios of several
emission lines.  Galaxies whose emission lines are powered by the
ionizing radiation field of young stars can be separated from those
powered by the harder radiation field produced by accretion onto a
massive black hole using an excitation diagram\cite{bpt81} (SI
Figure~\ref{fig:host}a).  The host galaxy of Swift\,J164449.3+573451
is clearly located in the portion of the diagram associated with
star-forming galaxies and not AGN.  A small systematic uncertainty
results from our inability to adequately correct for the underlying
stellar absorption features with the signal-to-noise ratio of our
current spectra, but that effect will only move the point associated
with Swift\,J164449.3+573451 down and to the left, closer to the main
locus of star-forming galaxies.  In addition, neither of our spectra
show any evidence for emission from high-excitation lines such as
${\rm [NeV]}\lambda\lambda 3345,3425$, which would be evidence for an
AGN.  In summary, the spectra of the host galaxy of
Swift\,J164449.3+573451 are consistent with expectations for a galaxy
whose central black hole was not actively accreting prior to the
current outburst.

\section{Synchrotron Model of the Radio Emission}
\label{sec:synch}

We assume that the radiation observed from Swift\,J164449.3+573451 in
the radio band corresponds to synchrotron emission from relativistic
electrons in a spherically expanding source.  We use as inputs the
observations at $\Delta t\approx 5$, 10, 15, and 22~d (see Figure~2 of
the main text).  Given the shape of the spectrum we use $\nu_a\approx
\nu_p$ in all three epochs.

We use the methods described in \S3 of Ref.~\pcite{kn09} to analyze
each epochs independently.  Although that work deals with gamma-ray
bursts, which are highly relativistic, the basic framework is quite
general.  We describe the source by means of five parameters: the
source radius, $r$, the bulk Lorentz factor, $\Gamma$, the Lorentz
factor of the relativistic electrons that produce the radiation at the
synchrotron peak, $\gamma_e$ (in the source frame), the number of
relativistic electrons, $N_e$, and the magnetic field strength $B$ (in
the source frame).  To determine these five parameters, we need five
constraints.  Three of these are provided by the measured values of
$\nu_p$, $F_{\nu,p}$, and $\nu_a$, coupled with equations 1, 2, and
15 in Ref.~\pcite{kn09}.

A fourth condition is obtained by the assumption that the magnetic
field and the relativistic particles are in equipartition.  We write
this condition as (see equations 33 and 34 Ref.~\pcite{kn09}):
\begin{equation}
E_B = B^2 r^3/2 = 10\,E_e = 10\,N_e\gamma_e\Gamma m_ec^2, 
\label{RN1}
\end{equation}
where the factor of 10 assumes that the total thermal energy in both
electrons and protons is 10 times that in electrons alone. 

Finally, for the fifth condition, we require the radius of the source
to be consistent with ballistic expansion from the moment of the
initial burst.  The observed time of $\Delta t=5$~d corresponds to a
time of 3.7~d in the frame.  Due to relativistic time compression, the
actual time that the source would have expanded is $(3.7\,{\rm
d})/(1-\beta)$, where $\beta=v/c$ is the relativistic bulk expansion
velocity of the source.  Thus, the assumption of ballistic motion
requires:
\begin{equation}
r = (3.7\,{\rm d})\times \beta c/(1-\beta). 
\label{RN2}
\end{equation}

Using the above five conditions, we obtain solution for the relevant
parameters of the source at all four epochs.  These solutions are
summarized in Table~\ref{tab:model}.  We find that all four epochs
lead to the same expansion velocity of $\Gamma\approx 1.2$ (i.e., no
deceleration) and an energy of $E_B = 10\,E_e\approx 1.6\times
10^{50}$ erg.  The outflow mass, $8\times 10^{-5}$ M$_\odot$, combined
with the constant expansion velocity, indicate that the density of the
swept-up medium is $\simlt 3\times 10^3$~cm$^{-3}$.

\begin{figure}
\centerline{\psfig{file=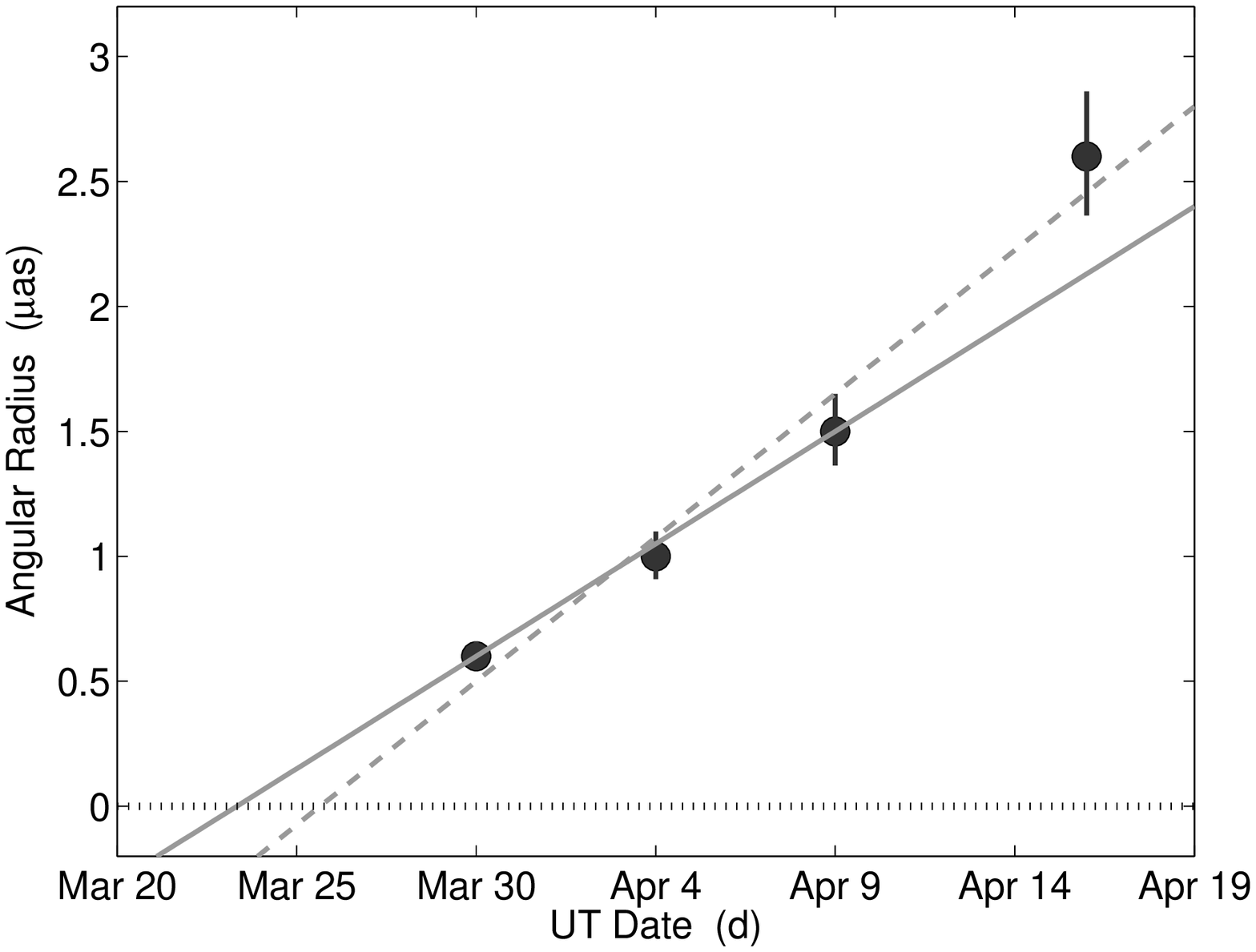,width=6.5in,angle=0}}
\caption[]{\small The radius of the radio emitting region inferred
from the relativistic model indicates a formation date in excellent 
agreement with the intial burst.  Angular radius of the
radio emitting region as a function of UT date.  The linear
growth in source size indicates a formation date of 2011 March 23-26
UT, in agreement with the initial {\it Swift}
detection of $\gamma$-ray emission on 2011 March 25 UT.  The solid and
dashed lines indicate fits to the first three epochs and all four
epochs, respectively.  We note that the last epoch on 2011 April 16 UT
has a larger uncertainty than the preceding measurements.}
\label{fig:vel} 
\end{figure}

Fitting a linear trend to the expanding source size (as indicated by
the data), we can estimate the initial formation date of the
relativistic outflow (Figure~\ref{fig:vel}).  We find that the this
date is 2011 March $23-26$ UT, in excellent agreement with
the initial {\it Swift} detection of $\gamma$-ray emission on 2011
March 25 UT.  This indicates that the formation of the relativistic
outflow indeed coincided with the initial accretion episode onto the
supermassive black hole.

Using the model, we have computed the synchrotron self-Compton (SSC)
emission and find that the peak occurs at $\approx 0.1$ keV and the
luminosity at the peak is $\approx 2\times 10^{45}$~erg~s$^{-1}$.
These values do not agree with the observed X-ray spectrum which peaks
beyond 1 keV and has a peak luminosity of $\approx 10^{47}$~erg~s$^{-1}$ 
(observer frame).  Thus, the model developed here suggests
that the X-rays must arise either from a process other than SSC (e.g.,
Compton scattering of external radiation) or from a different region
of the source.

Interestingly, in the framework of this model relativistic expansion
with $\Gamma\sim 10$, as seen in most blazars, is ruled out.  For
$\Gamma\sim 10$, the radius $r$ required to fit the observed radio
flux and spectrum is much smaller than the radius implied by ballistic
dynamics.  It may be possible to avoid this difficulty by modeling the
source as a highly collimated jet.  However, the collimation angle of
the jet would need to be much smaller than $1/\Gamma$, which is
unlikely to be true.  Another approach is to give up the condition of
energy equipartition.

\clearpage
\begin{center}
\begin{longtable}{lcccc}
\caption{Summary of Radio Observations} 
\label{tab:radio} \\
\hline
\hline
UT Date & $\Delta t$ & Facility & $\nu$ & $F_\nu$ \\
        & (d)        &          & (GHz) & (mJy)   \\\hline
\endfirsthead
\hline
\hline
UT Date & $\Delta t$ & Facility & $\nu$ & $F_\nu$ \\
        & (d)        &          & (GHz) & (mJy)   \\\hline
\endfirsthead
\endhead
Mar 31.28   &  2.74  & EVLA & 1.4 & $<0.30$  \\
Apr 1.28    &  3.74  & EVLA & 1.4 & $<0.18$  \\\hline
Mar 31.28   &  2.74  & EVLA & 1.8 & $<0.15$  \\
Apr 1.28    &  3.74  & EVLA & 1.8 & $<0.15$  \\\hline
Mar 29.36   &  0.82  & EVLA & 4.9 & $0.25\pm 0.01$  \\
Mar 30.25   &  1.71  & EVLA & 4.9 & $0.34\pm 0.02$  \\
Mar 30.49   &  1.95  & EVLA & 4.9 & $0.34\pm 0.02$  \\
Mar 31.28   &  2.74  & EVLA & 4.9 & $0.61\pm 0.02$  \\
Apr 1.27    &  3.73  & EVLA & 4.9 & $0.82\pm 0.02$  \\
Apr 2.26    &  4.72  & EVLA & 4.9 & $1.48\pm 0.02$  \\
Apr 4.28    &  6.74  & EVLA & 4.9 & $1.47\pm 0.02$  \\
Apr 9.47    &  11.93 & EVLA & 4.9 & $1.80\pm 0.03$  \\
Apr 17.27   &  19.73 & EVLA & 4.9 & $2.11\pm 0.01$  \\\hline
Mar 29.36   &  0.82  & EVLA & 6.7 & $0.38\pm 0.01$  \\
Mar 30.25   &  1.71  & EVLA & 6.7 & $0.63\pm 0.02$  \\
Mar 30.49   &  1.95  & EVLA & 6.7 & $0.64\pm 0.02$  \\
Mar 31.28   &  2.74  & EVLA & 6.7 & $1.16\pm 0.02$  \\
Apr 1.27    &  3.73  & EVLA & 6.7 & $1.47\pm 0.02$  \\
Apr 2.26    &  4.72  & EVLA & 6.7 & $1.50\pm 0.02$  \\
Apr 4.28    &  6.74  & EVLA & 6.7 & $2.15\pm 0.02$  \\
Apr 9.47    &  11.93 & EVLA & 6.7 & $3.79\pm 0.03$  \\
Apr 17.27   &  19.73 & EVLA & 6.7 & $3.44\pm 0.01$  \\\hline
Apr 9.46    &  11.92  & EVLA & 8.4 & $5.49\pm 0.09$  \\\hline
Mar 31.33   &  2.79  & AMI-LA & 15.0 & $2.80\pm 0.45$ \\
Apr 1.18    &  3.65  & AMI-LA & 15.0 & $3.58\pm 0.23$ \\
Apr 2.16    &  4.62  & AMI-LA & 15.0 & $4.35\pm 0.28$ \\
Apr 3.10    &  5.56  & AMI-LA & 15.0 & $5.10\pm 0.36$ \\
Apr 4.10    &  6.56  & AMI-LA & 15.0 & $6.67\pm 0.51$ \\
Apr 5.33    &  7.79  & AMI-LA & 15.0 & $6.65\pm 0.48$ \\
Apr 6.10    &  8.56  & AMI-LA & 15.0 & $7.36\pm 0.45$ \\
Apr 8.03    &  10.49 & AMI-LA & 15.0 & $8.30\pm 0.24$ \\
Apr 8.39    &  10.85 & AMI-LA & 15.0 & $7.01\pm 0.13$ \\
Apr 9.12    &  11.58 & AMI-LA & 15.0 & $8.50\pm 0.32$ \\
Apr 13.20   &  15.66 & AMI-LA & 15.0 & $8.66\pm 0.40$ \\
Apr 14.27   &  16.73 & AMI-LA & 15.0 & $9.91\pm 0.63$ \\
Apr 16.33   &  18.79 & AMI-LA & 15.0 & $10.96\pm 0.89$ \\
Apr 17.30   &  19.76 & AMI-LA & 15.0 & $10.78\pm 0.77$ \\\hline
Mar 31.37   & 2.82  & OVRO40 & 15.0 & $2.36\pm 1.14$ \\
Apr 1.34    & 3.80  & OVRO40 & 15.0 & $3.17\pm 1.17$ \\
Apr 8.36    & 10.82 & OVRO40 & 15.0 & $8.06\pm 0.95$ \\
Apr 10.36   & 12.82 & OVRO40 & 15.0 & $7.65\pm 0.90$ \\
Apr 12.33   & 14.79 & OVRO40 & 15.0 & $7.45\pm 0.80$ \\
Apr 12.41   & 14.87 & OVRO40 & 15.0 & $6.60\pm 1.22$ \\
Apr 15.35   & 17.81 & OVRO40 & 15.0 & $10.54\pm 0.93$ \\
Apr 19.33   & 21.79 & OVRO40 & 15.0 & $11.99\pm 1.03$ \\
Apr 21.36   & 23.82 & OVRO40 & 15.0 & $9.55\pm 1.04$ \\\hline
Mar 30.24   & 1.70 & EVLA & 19.1 & $2.12\pm 0.02$   \\
Apr 1.24    & 3.70 & EVLA & 19.1 & $4.36\pm 0.05$   \\
Apr 2.26    & 4.72 & EVLA & 19.1 & $5.25\pm 0.03$   \\
Apr 3.36    & 5.82 & EVLA & 19.1 & $6.38\pm 0.05$   \\
Apr 4.27    & 6.73 & EVLA & 19.1 & $5.31\pm 0.03$   \\
Apr 16.38   & 18.84 & EVLA & 19.1 & $12.01\pm 0.03$  \\\hline
Mar 30.24   & 1.70 & EVLA & 24.4 & $3.01\pm 0.03$   \\
Apr 1.24    & 3.70 & EVLA & 24.4 & $5.58\pm 0.06$   \\
Apr 2.26    & 4.72 & EVLA & 24.4 & $6.70\pm 0.03$   \\
Apr 3.36    & 5.82 & EVLA & 24.4 & $7.88\pm 0.12$   \\
Apr 4.27    & 6.73 & EVLA & 24.4 & $6.60\pm 0.03$   \\
Apr 16.38   & 18.84 & EVLA & 24.4 & $12.69\pm 0.03$  \\\hline
Mar 31.28   & 2.70 & EVLA & 44 & $6.35\pm 0.10$  \\
Apr 1.24    & 3.70 & EVLA & 44 & $7.71\pm 0.08$  \\
Apr 3.36    & 5.82 & EVLA & 44 & $9.38\pm 0.08$  \\\hline
Apr 2.68    & 5.14  & CARMA & 87 & $18.6\pm 0.3$   \\
Apr 3.63    & 6.09  & CARMA & 87 & $21.7\pm 0.2$   \\
Apr 4.72    & 7.18  & CARMA & 87 & $14.6\pm 0.4$   \\
Apr 6.63    & 9.09  & CARMA & 87 & $15.1\pm 0.2$   \\
Apr 12.15   & 14.61 & CARMA & 87 & $10.4\pm 0.2$   \\
Apr 16.60   & 19.06 & CARMA & 87 & $9.36\pm 0.5$   \\
Apr 19.61   & 22.07 & CARMA & 87 & $5.49\pm 0.3$   \\\hline
Mar 30.39   & 1.85  & CARMA & 94 & $15.7\pm 0.3$   \\
Apr 16.60   & 19.06 & CARMA & 94 & $10.7\pm 1.0$   \\\hline
Apr 13.66   & 16.12 & SMA & 200 & $14.1\pm 1.5$    \\
Apr 18.70   & 21.16 & SMA & 200 & $8.2\pm 1.2$     \\
Apr 20.74   & 23.20 & SMA & 200 & $7.4\pm 1.0$     \\\hline
Mar 28.82   & 0.28  & SMA & 225 & $<33$            \\
Apr 4.74    & 7.20  & SMA & 225 & $14.9\pm 1.5$    \\
Apr 5.40    & 7.86  & SMA & 225 & $11.7\pm 0.4$    \\
Apr 11.63   & 14.10 & SMA & 225 & $13.3\pm 1.5$    \\
Apr 12.66   & 15.12 & SMA & 225 & $9.9 \pm 1.4$    \\
Apr 14.65   & 17.11 & SMA & 225 & $8.2 \pm 1.4$    \\
Apr 15.67   & 18.13 & SMA & 225 & $8.3 \pm 2.2$    \\\hline
Mar 30.39   & 1.85 & SMA & 345   & $35.1\pm 1.0$  \\
Mar 31.39   & 1.85 & SMA & 345   & $<10.2$        \\
Apr 1.71    & 4.17 & SMA & 345   & $<8.7$         \\\hline
\caption[]{Radio observations of Swift\,J164449.3+573451.  All values
of $\Delta t$ are relative to the $\gamma$-ray trigger.  For times
relative to the initial detection (2011 March 25 UT) as used in the
main text add 3.54 d.  Flux measurement errors are one standard
deviation systematic errors.  Absolute flux scaling depends on the
instrument.  See SI text for details.}
\label{tab:radio}
\end{longtable}
\end{center}

\clearpage
\begin{center}
\begin{longtable}{lcccc}
\caption{Summary of Radio SED Modeling} \\
\hline
\hline
Parameter & $\Delta t=5$ d & $\Delta t=10$ d & $\Delta t=15$ d  & $\Delta t=22$ d \\
\hline
\hline
$\nu_a=\nu_p$ (GHz; rest-frame)  & 600 & 250 & 140 & 80   \\
$F_{\nu,p}$ (mJy; rest-frame)    & 80  & 40  & 30  & 25   \\\hline 
$r$ ($10^{16}$ cm)               & 1.0 & 1.7 & 2.6 & 4.7  \\
$\Gamma$                         & 1.2 & 1.2 & 1.1 & 1.2     \\
$\beta$                          & 0.5 & 0.5 & 0.5 & 0.5      \\
$\gamma_e$                       & 150 & 140 & 140 & 140      \\
$B$ (G)                          & 17  & 8   & 4.6 & 2.6      \\
$N_e$ ($10^{53}$)                & 1.0 & 1.1 & 1.5 & 2.6      \\
$n_e$ (10$^4$ cm$^{-3}$)       & 2.4 & 0.5 & 0.2 & 0.06     \\
$E_B=10E_e$ ($10^{50}$ erg)      & 1.4 & 1.5 & 1.9 & 2.9      \\
$\theta_s$ ($\mu$as)             & 0.6 & 1.0 & 1.5 & 2.6      \\\hline 
\caption[]{Summary of relativistic model results for the four
broad-band SEDs shown in Figure 2 of the main text.  The top portion
lists the observed synchrotron parameters, while the bottom portion
lists the model fit results.}
\label{tab:model}
\end{longtable}
\end{center}

\clearpage

\clearpage
